# Influence of orbital symmetry on diffraction imaging with rescattering electron wave packets

M. G. Pullen[1], B. Wolter[1], A.-T. Le[2], M. Baudisch[1], M. Sclafani[1], H. Pires[1], C. D. Schröter[3], J. Ullrich[3,4], R. Moshammer[3], T. Pfeifer[3], C. D. Lin[2], J. Biegert[1,5]

[1]ICFO - Institut de Ciencies Fotoniques, The Barcelona Institute of Science and Technology, 08860 Castelldefels (Barcelona), Spain
[2]J.R. Macdonald Laboratory, Physics Department, Kansas State University, Manhattan, Kansas 66506-2604, USA
[3]Max-Planck-Institut für Kernphysik, Saupfercheckweg 1, 69117 Heidelberg, Germany
[4]Physikalisch-Technische Bundesanstalt, Bundesallee 100, 38116 Braunschweig, Germany
[5]ICREA - Institució Catalana de Recerca i Estudis Avançats, 08010 Barcelona, Spain
e-mail: jens.biegert@icfo.eu

**The ability to directly follow and time resolve the rearrangement of the nuclei within molecules is a frontier of science that requires atomic spatial and few-femtosecond temporal resolutions. While laser induced electron diffraction can meet these requirements, it was recently concluded that molecules with particular orbital symmetries (such as $\pi_g$) cannot be imaged using purely backscattering electron wave packets without molecular alignment. Here, we demonstrate, in direct contradiction to these findings, that the orientation and shape of molecular orbitals presents no impediment for retrieving molecular structure with adequate sampling of the momentum transfer space. We overcome previous issues by showcasing retrieval of the structure of randomly oriented $O_2$ and $C_2H_2$ molecules, with $\pi_g$ and $\pi_u$ symmetries, respectively, and where their ionisation probabilities do not maximise along their molecular axes. While this removes a serious bottleneck for laser induced diffraction imaging, we find unexpectedly strong back scattering contributions from low-Z atoms.**

Knowledge about molecular structure is crucial to the understanding of complex chemical and biological systems. Gas-phase diffraction techniques [1] provide well established standards to image static structure. However, a deeper understanding of a system and its function only comes with temporally resolved measurements that enable the nature and timescales of how different components interact with each other to be observed. The advent of intense and ultrashort laser pulses prompted the development of techniques such as ultrafast electron diffraction [2, 3] and X-ray diffraction [4], which are nowadays tools to study molecular dynamics on previously inaccessible temporal scales approaching a few hundred femtoseconds. However, to fully capture the entire dynamics of structural rearrangement, one needs to access the triggering events of a transformation, which demands access to the few femtosecond temporal range. Not surprisingly, tremendous efforts are being made to access these timescales with current and emerging techniques [5-8]. Laser-induced electron diffraction (LIED) [9] is a maturing technique that has already achieved few-femtosecond resolution on homonuclear diatomics [10] and it has recently been extended to polyatomic molecules [11].

In LIED, the momentum distribution of strong-field induced rescattered electrons is analysed as a function of the incident return momentum ($k_r$) and scattering angle ($\vartheta_r$) [12] and spatial resolutions of 5 pm are achievable when combining mid-IR rescattering [13] with kinematically complete detection [11]. While LIED makes use of the doubly differential scattering cross section as a whole, a simpler analysis was recently presented by Xu *et al.* [14] in which only backscattered ($\vartheta_r \approx 180°$) electron distributions are analysed. The appeal of using only a 1D data set instead of the 3D doubly differential scattering cross section lies in the faster data extraction procedure without the need for iterative algorithms. Moreover, such a Fourier analysis based 1D version of LIED, FT-LIED, achieves much faster structural retrieval in a manner that is similar to ultrafast electron diffraction. However, when investigating the same simple diatomic species as in Ref. [10], Xu *et al.* found that FT-LIED succeeded in retrieving the structure of $N_2$, but failed for $O_2$ due to a lack of structural information encoding in the respective backscattered electron distribution. They identified the failure of the FT-LIED methodology as being related with the ionisation probability of the target molecule, which maximises parallel with the bond axis for $N_2$ but not for $O_2$. From their findings, and based on their theoretical investigation, the authors concluded that in order to image molecular bonds with the FT-LIED technique the internuclear axis of interest must be aligned with the laser polarisation direction. This conclusion has profound implications for LIED as it presents a strong impediment for further advancement of FT-LIED since most larger and complex molecules are not readily aligned (or oriented), and even those that can be would apparently require time-consuming tomographic techniques.

Here, we address the issue of the influence of the orbital symmetry of the target molecule and establish that it does not impede the retrieval of molecular structure. To demonstrate the utility of FT-LIED, we successfully image the structure of both $O_2$ and of $C_2H_2$ without molecular alignment and irrespective of the fact that molecular ionisation is minimal along the molecular axis for both target species. Our results clearly show that the ionisation angular dependence of the highest occupied molecular orbital (HOMO) does not prevent imaging of internuclear distances. This demonstrates that imaging is achievable from isotropic gas samples as routinely observed in ultrafast electron diffraction. We do, however, find that molecular structure can be obscured when the momentum transfer space is inadequately sampled. In addition, in our analysis we find unexpectedly strong scattering from hydrogen atoms. This observation presents a noticeable departure from the theoretical foundation of diffraction imaging, the independent atom model, which underestimates the contribution for the scattering conditions used here and warrants further theoretical investigations for the collision energies accessible to LIED.

**Results**

**Randomly oriented molecules and retrieval of their structure with LIED.** We first illustrate that the angle of maximum ionisation from the HOMO does not need to coincide with the laser polarisation axis to retrieve molecular structure. We choose as examples $O_2$ (blue, solid throughout Fig. 1) and $C_2H_2$ (red, dashed throughout Fig. 1), whose HOMOs possess $\pi_g$ and $\pi_u$ symmetry, respectively, and in neither case do their ionisation probabilities maximise parallel to the laser polarisation direction (i.e. at 0° or 180°), as shown in Fig. 1a. Electrons are predominantly emitted at 90° and 270° for $C_2H_2$ and at angles ~45° away from parallel for $O_2$. In order not to limit our investigation by the detection method, we resort to 3D electron-ion coincidence detection with a reaction microscope (see methods for details) to measure the backscattered electron distributions. We have previously shown [5] that our instrument's coincidence capabilities are important to provide high signal-to-noise data by discriminating backscattered electron distributions from different ionic species that are produced when ionizing the molecular targets. For a simple diatomic this issue is not necessarily a problem as only a few ionic components contribute. In the case of larger and more complex molecules, however, the number of detected ionic fragments will be large and their yields can be comparable to or greater than the main ion [15].

Figure 1b shows the interference signals measured for $O_2^+$ and $C_2H_2^+$ as a function of the momentum transfer experienced by the diffracted electrons ($q = 2k_r\sin(\vartheta_r/2)$). We have isolated the molecular modulations from the experimental backscattered electron distributions by implementing empirical background subtraction (details are given in the Supplementary Information in section "background subtraction"), in a manner analogous to ultrafast electron diffraction [1]. This method presents a number of important advantages compared with previous implementations of LIED. Firstly, the structure of simple molecules can be determined without depending on theoretical calculations. Secondly, experimental data acquisition times will be reduced significantly as the backscattered electron distributions from partner atoms do not need to be measured. Thirdly, and most importantly, all molecules are now potential targets as even those without partner atoms can be investigated. After background subtraction, is it observed that the resultant interference signals posses a number of extrema and zero-crossings, which are sensitive to molecular structure at the time of rescattering. The phase, frequency and amplitude are observed to be molecule dependent, which is a first indication that structural information is present. We note that, in general, the modulation amplitude of the experimental backscattered electron distributions are greater than those predicted by the independent atom model.

Fourier transforming the measured interference signals results in a radial distribution spectrum that contains frequency components at the molecular internuclear distances. Before transformation a windowing function and zero padding are applied to the experimental data. The normalised spectra resulting from the Fourier transform of the interference signals in Fig. 1b are presented in Fig. 1c. Also indicated are the expected positions of the $O_2$ and $C_2H_2$ cation internuclear distances [16]. The inset shows the region between 0.9 - 1.5 Å where the anticipated O-O and C-C bond lengths of 1.12 Å and 1.25 Å, respectively, are located. The experimental peak positions are within 0.05 Å of these values. This finding clearly confirms that the bonds of molecules do not need to be aligned with the angle of maximum ionisation rate in order to be successfully imaged.

We find that the main $C_2H_2$ cation peak near 1.3 Å is likely composed of two internuclear distances that are within <0.2 Å of each other and therefore not individually resolvable: the C-C bond mentioned above at 1.25 Å and the C-H (and H-C) bond at 1.08 Å (see inset of Fig. 1c). Due to the higher cross-section of C compared to H it is expected and observed that this peak should be shifted more towards the position of C-C than C-H. Interestingly and unexpectedly, a second and a third peak are also visible at distances of 2.33 Å and 3.51 Å and amplitudes of 0.40 and 0.11, respectively. The position of the second peak agrees with the cation C-C-H and H-C-C internuclear distance of 2.34 Å. This shows that scattering from the H atom is not negligible as expected from the independent atom model. We discard the third peak of the $C_2H_2$ spectrum due to low signal to noise even though it contains a signal corresponding to the expected H-C-C-H position. The exclusion of this region is supported by the fact that an artefact is found for $O_2$ with similarly low amplitude at similar position (3.21 Å).

**Limitations of the independent atom model.** The experimental diffraction data permitted to extract, for both species, $O_2$ and $C_2H_2$, the most prominent contributions of heavy atoms corresponding to O-O and C-C distances, respectively. It is interesting, however, that we obtain clear signatures corresponding to the C-H and C-C-H bond distances. These experimental findings contradict calculations from the independent atom model, which predict that the differential cross-section (DCS) for H atoms ($\sigma_H$) should be one order of magnitude lower than for C atoms ($\sigma_C$) for the typical scattering parameters used in LIED ($E_r$ = 20-240 eV and $\vartheta_r$ = 180°). We compare our results to the independent atom model by calculating the expected DCS as a function of returning electron energy at a constant scattering angle of $\vartheta_r$ = 180°. The result of the calculation for neutral $C_2H_2$ (grey, solid in all panels) is presented alongside our experimental results (red, dashed in all panels) in Fig. 2a. The independent atom model curve is dominated by one frequency component only, which can be observed at a position near 1.21 Å (neutral C-C bond length) after Fourier transformation. Thus theory suggests that the influence of H atoms should not be observable in FT-LIED experiments for our experimentally accessible parameters. Note that the slight difference in the position of the main peak is due to the fact that neutral $C_2H_2$ is used in the simulation while the cation is measured in the experiment. If the independent atom model calculation is artificially modified by increasing the $\sigma_H/\sigma_C$ ratio by a factor of five then other frequency components start to contribute to the interference signal. The most interesting finding is that the Fourier transformed spectrum (Fig. 2b) now closely resembles the experimental one with a new contribution near 2.3 Å that is similar in magnitude to the measured one. Upon further increasing the $\sigma_H/\sigma_C$ ratio to a factor of ten, three Fourier peaks are visible and the similarity between the two curves clearly shows that the influence of H scattering can be observed in the experimental data (Fig. 2c). The amplitude of the 2.33 Å peak is about 40% of the main peak and therefore cannot be discarded. Its position is equal to the sum of the retrieved C-C plus C-H bond lengths therefore corresponding to the distance of C-C-H. Moreover, the smaller Fourier amplitude of C-C-H compared with C-C is consistent with a larger separation thus indicating that the C-C-H structure was successfully retrieved. We may use the identical arguments to state that the small peak near 3.5 Å is the interference from the two H atoms. Based on the current independent atom model (Fig. 2a) it is hard to assess whether the experimental peak near 3.5 Å is related to molecular structure.

Our results suggest that the independent atom model does not fully describe electron scattering for the collision energies that are typical for FT-LIED. We note that differences between $C_2H_2$ independent atom model calculations and electron impact DCS measurements have also been observed previously for similar scattering energies [17]. Using the entire doubly differential cross section for the data analysis, the IAM has shown excellent agreement with experimental $C_2H_2$ LIED data, however, this was at smaller scattering angles than 180 degrees [11]. While the apparent failure of the IAM is an important issue that should stimulate further investigations, both theoretical and experimental, our results show that the FT-LIED method continues to work irrespective of this failure. The larger relative contribution of hydrogen with respect to heavier atoms may become an advantage of LIED for studying dynamics involving proton migration, which plays a major role in many chemical reactions.

**Importance of sampling the momentum transfer space.** We now address why the structure of $O_2$ could not be imaged in Ref. [14]. In contrast with ultrafast electron diffraction, LIED makes use of the target molecule's own electrons for imaging. The LIED mechanism is based on strong-field tunnel ionisation of the target molecule and rescattering off the target molecular ion. Field-free elastic scattering cross sections are extracted by invoking the strong-field rescattering model, which can be fulfilled in the quasistatic, or deep tunnelling, regime. The unavoidable consequence of using a strong-field to image molecules that possess relatively low ionisation potentials (12.07 eV for $O_2$ [18]) is the resulting number of strong-field processes such as fragmentation, Coulomb explosion and multiple ionisation. Each of these processes has associated rescattering electrons carrying different momenta that obstruct the modulations observed in molecular backscattered electron distributions and therefore prohibit structural retrieval.

To corroborate our explanation we illustrate the influence of these adverse processes on the image bearing backscattered electron distributions. Figure 3a shows ion time-of-flight data after ionisation of $O_2$ from which we had extracted backscattered electron distributions which were only associated to $O_2^+$ (Fig. 1b). Now we illustrate the importance of our selection of the momentum transfer space by eliminating coincidence conditions. A number of positively charged particles are detected including the $O_2^+$ ion located near 5.5 µs (blue shading in inset). In Fig. 3b we compare the backscattered electron distributions extracted when using electrons coincident with $O_2^+$ only (blue, dashed - our conditions) and with all detected ions (orange, solid). Also shown are fourth order polynomial fits to the backscattered electron distributions (black, dotted). Two observations are evident: 1) an order of magnitude difference between the two backscattered electron distributions is visible over a large range of the spectrum, indicating that there are many contaminating electrons even for a simple species such as $O_2$, and 2) the superimposed modulation is significantly enhanced when these contaminating electrons are excluded. The factor of three difference in modulation is more discernible after subtraction of the empirical background, as presented in Fig. 3c. In this case the signals are displayed with the windowing function already applied. Single sinusoidal fits to the data (grey, dotted) display how well the observed modulations can be represented by a single frequency and also help to quantify the decreased modulation.

The results in Fig. 3 show that even when using a relatively moderate laser intensity of $8.5 \times 10^{13}$ W cm$^{-2}$, a significant unwanted electron signal is present during mid-IR strong-field ionisation of molecular targets. In the case of $O_2$ the desired electrons cannot be discriminated from this omnipresent unwanted signal without coincidence selection of the momentum transfer space. The peak laser intensity used to investigate $O_2$ in Ref. [14] was over 50% greater than we use here.

According to molecular Ammosov-Delone-Krainov simulations [19] this would translate to an order of magnitude higher ionisation rate. Thus, even more electrons would be generated from other strong-field processes and it is these electrons that prohibited structural imaging of $O_2$. In the case of a molecule such as $N_2$, which was successfully imaged, its higher ionisation potential of 15.58 eV [20] causes it to be harder to both ionise and fragment. Therefore the signal measured would most likely be dominated by $N_2^+$ electrons.

**Discussion**

We have investigated whether molecular orientation has a prohibitive influence on the usage of backscattering electron wavepackets to imaging molecular structure. Our investigation provides clear evidence that atomic scale diffraction imaging does not require alignment of the target species with FT-LIED. Moreover, we use a simple empirically background subtracted FT-LIED methodology with which we achieve imaging of two entirely different structures, the diatomic $O_2$ and polyatomic $C_2H_2$ molecules. Overall, our results have important consequences for FT-LIED and atomic scale imaging. Firstly, the HOMO structure of an unaligned molecule poses no limitation on diffraction imaging with backscattering wavepackets. Secondly, all molecules, rather than just those that can be aligned to a high degree, will be open to investigation. Lastly, we find that accurate bond lengths, particularly those involve atomic hydrogen, can still be accurately extracted using the FT-LIED method when the IAM begins to fail. This is an important advantage for FT-LIED that makes the investigation of time resolved studies in complex molecules involving proton migration or isomerisation possible. Further theoretical and experimental investigations with other hydrogen-rich molecules will be needed to firmly establish this important aspect of FT-LIED.

## Methods

**Light source.** Our mid-IR source is a home-built optical parametric chirped pulse amplification system that generates a number of wavelengths including a λ = 3.1 μm output that has a duration of 75 fs FWHM at a repetition rate of 160 kHz [21]. The high repetition rate more than compensates for the unfavourable $\lambda^{-4}$ scaling of strong-field induced rescattering [22]. A 50 mm focal length mirror focusses the 3.1 μm radiation to a spot size of about 6-7 μm, which results in a peak intensity of about I = 8.5 x $10^{13}$ W cm$^{-2}$. The intensity was estimated using a variety of techniques. Such a peak intensity results in a Keldysh parameter of ɣ ≈ 0.3 and a ponderomotive energy of $U_P$ = 75 eV, which corresponds to maximum classical return and backscattered electron energies of $E_{ret,max}$ = 3.17$U_P$ ≈ 240 eV and $E_{back,max}$ = 10$U_P$ ≈ 750 eV, respectively. Such high-energy photoelectrons are those that scatter near the nucleus of each atom where the potential from the core dominates the influence from the outer-shell electron distribution [23].

**Detection system.** The laser focus intersects with a rotationally cold gas jet (T ≤ 100 K for both molecules) that passes through two skimmer stages. Both gases are of high purity. The interaction takes place in an ultra-high vacuum chamber that has a base pressure $10^{-11}$ mbar without gas load and that houses a reaction microscope detection system [24]. The reaction microscope is able to detect all charged products of the interaction in full coincidence and with 3D momentum information. The magnetic and electric fields strengths (12.9 G and 550 V, respectively) were set such that no 'dark regions' (otherwise known as nodes) were located in the half of the electron momentum distribution used during data analysis. There are two ways in which the coincidence capabilities of the reaction microscope are of importance to this study. Firstly, the capacity to isolate the electrons of interest (i.e. those correlated to $O_2^+$ or $C_2H_2^+$ only) from those corresponding to other strong-field processes so that uncontaminated backscattered electron distributions can be obtained. Secondly, the ability to emulate a time-of-flight spectrometer by extracting backscattered electron distributions that include electronic contributions related to all ion fragments created during the experiment.

The data that support the findings of this study are available from the corresponding author upon request.

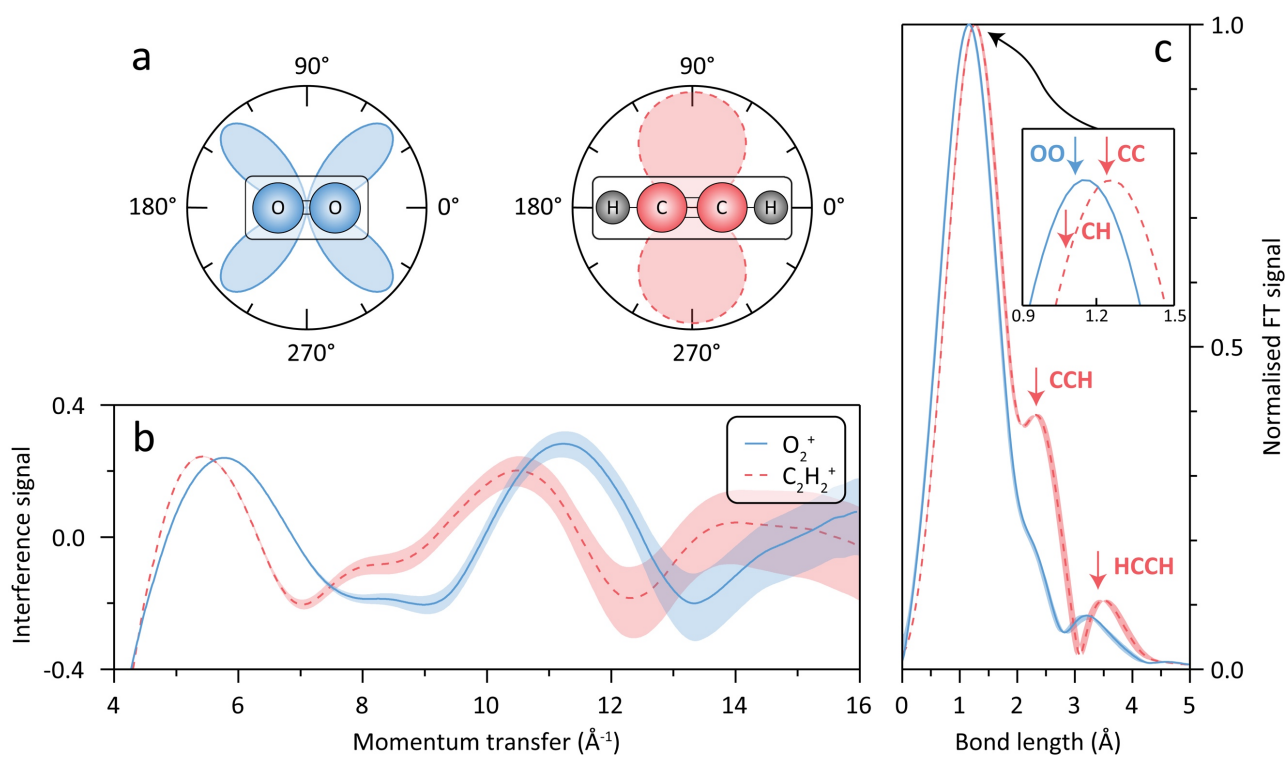

**Figure 1 Extracting bond lengths independent of HOMO structure.** (a) Simulated ionisation probabilities of the $O_2$ (blue solid curves throughout figure) and $C_2H_2$ (red dashed curves throughout figure) HOMOs as a function of the angle between the molecular axes and the laser polarisation direction. (b) The interference signals obtained for $O_2$ and $C_2H_2$ after ionisation of the molecules by the 3.1 μm source. In this panel the shading represents the error bars are estimated via Poissonian statistics. (c) The result of Fourier transforming the interference signals. The expected positions of the $O_2$ and $C_2H_2$ cation internuclear distances are indicated. The inset shows a zoomed in view around 1.2 Å where a difference of about 0.14 Å can be observed between the main peaks of the two molecules. The shaded regions represent the estimated error in the extracted spectra resulting from the uncertainty in the value of the ponderomotive energy.

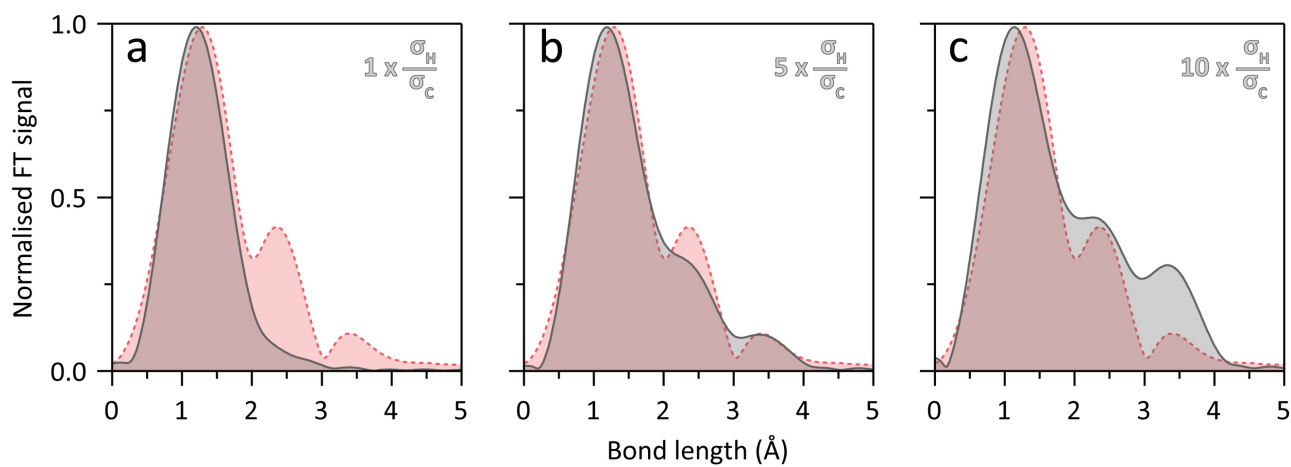

**Figure 2 Limitations of the independent atom model.** A comparison of the theoretical (grey, solid throughout figure) and experimental (red, dashed throughout figure) $C_2H_2^+$ Fourier spectra. In a the independent atom model is unmodified while in b the $\sigma_H/\sigma_C$ ratio is increased by a factor of five and in c by a factor of ten. All spectra have been normalised to their maximum values.

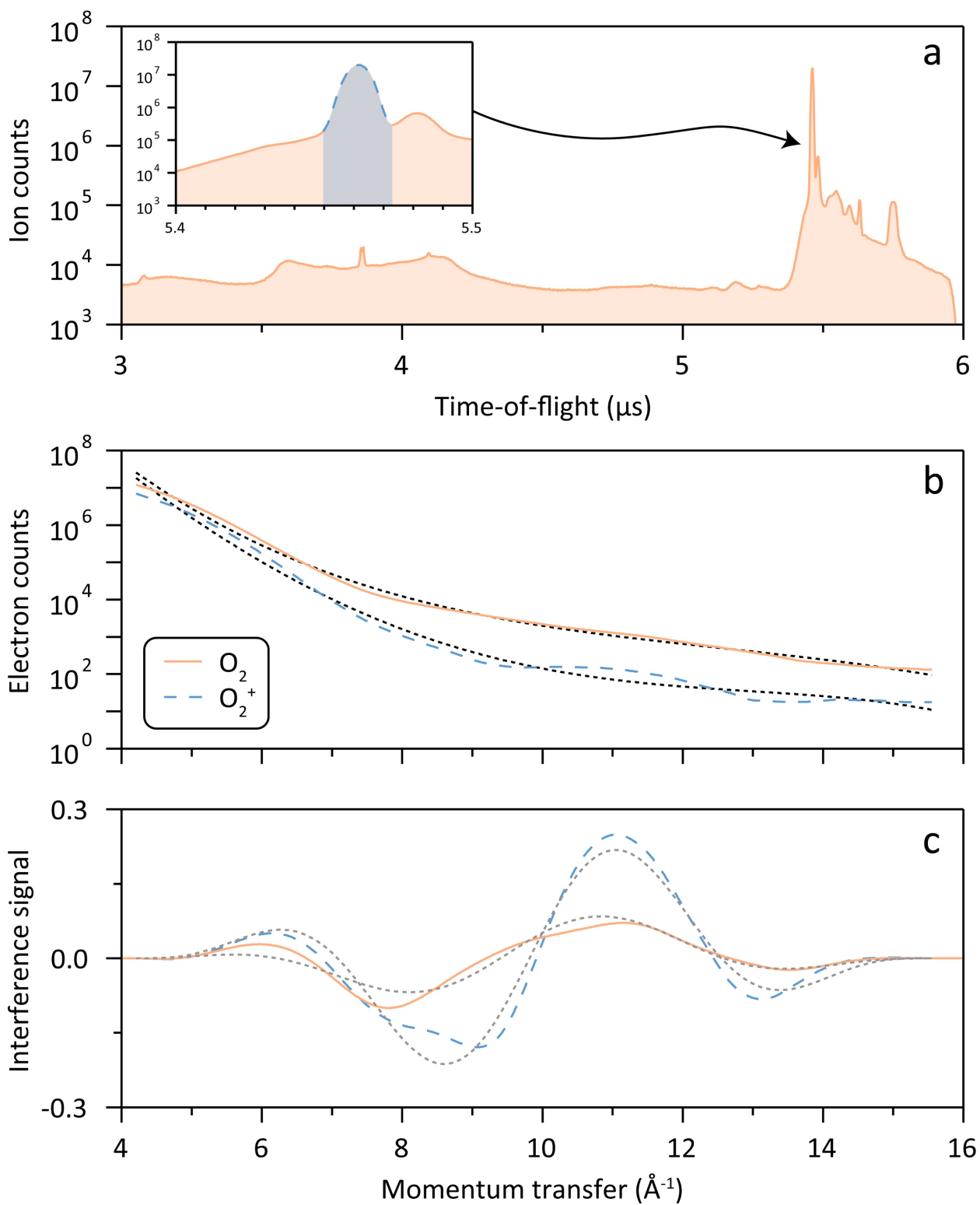

**Figure 3 Importance of correctly sampling the momentum transfer space**. (a) The ionic time-of-flight measured after the ionisation of $O_2$. The inset shows the small temporal range over which the $O_2^+$ ions are detected (blue shading). Other ions that aren't of interest in this manuscript are also detected (orange shading). (b) The extracted backscattered electron distributions when electrons coincident with all ions (orange, solid) or only $O_2^+$ (blue, dashed) are utilised. Fourth order polynomial fits are also presented (black, dotted). (c) The corresponding interference signals for $O_2^+$ (blue, dashed) and all (orange, solid) electrons. Single sinusoidal fits (grey, dotted) to the data show that the observable modulation has decreased by a factor of three.

**Acknowledgements**

We acknowledge financial support from the Spanish Ministry of Economy and Competitiveness, through FIS2014-56774-R, FIS2014-51478-ERC, the "Severo Ochoa" Programme for Centres of Excellence in R&D (SEV-2015-0522), the Catalan Agencia de Gestió d'Ajuts Universitaris i de Recerca (AGAUR) with SGR 2014-2016, Fundació Cellex Barcelona, the European Union's Horizon 2020 research and innovation programme under grant agreement No 654148 Laserlab-Europe, the Marie Sklodowska-Curie grant agreements No. 641272 and 264666, COST Action MP1203 and COST Action XLIC. M.G.P. is supported by the ICFONEST programme, partially funded by COFUND (FP7-PEOPLE-2013- COFUND) and B.W. was supported by AGAUR with a PhD fellowship (FI-DGR 2013–2015). A.-T.L. and C.D.L. are supported by the Chemical Sciences, Geosciences, and Biosciences Division, Office of Basic Energy Sciences, Office of Science, U.S. Department of Energy (DOE) under Grant No. DE-FG02-86ER13491.

**Author Contributions**

J. B. conceived the experiment. M. G. P., B. W., M. B., M. S. and H. P. performed the experiment. M. G. P., B. W. and M. S. performed the data analysis. A-. T. L. and C. D. L. provided theoretical support. C. D. S., J. U., R. M. and T. P. provided experimental support. M. G. P. and J. B. wrote the manuscript.

**Additional Information**

**Competing financial interests**: The authors declare that they have no competing financial interests.